\documentclass[aps,twocolumn,superscriptaddress,prc,showpacs]{revtex4}
\usepackage{psfig}
\usepackage{longtable}
\def\tj #1 #2 #3 #4 #5 #6 {\left( \begin{array}{ccc} #1 & #2 & #3 \\
#4 & #5 & #6 \end{array}\right) }
\begin{document}
\title{Magnetic moments of the 2${^+_1}$ states around $^{132}$Sn}
\author{B.~A.~Brown}
\affiliation{Department of Physics and Astronomy and National Superconducting
Cyclotron Laboratory, Michigan State University, East Lansing, MI 48824, USA}
\author{N.~J.~Stone}
\affiliation{Department of Physics,
University of Oxford, Oxford, OX1 3PU, United Kingdom}
\affiliation{Department of Physics and Astronomy, University of Tennessee, Knoxville, TN 37996, USA}
\author{J.~R.~Stone}
\affiliation{Department of Physics,
University of Oxford, Oxford, OX1 3PU, United Kingdom}
\affiliation{Department of Chemistry and Biochemistry, University of
Maryland, College Park, MD 20742, USA}
\author{I.~S.~Towner}
\affiliation{ Physics Department, Queen's University, Kingston, Ontario, K7L
3N6, Canada}
\author{M.~Hjorth-Jensen} 
\affiliation{Department of Physics and Center of Mathematics for Applications, University of Oslo, N-0316, Oslo, Norway}
\affiliation{PH Division, CERN, CH-1211 Geneva 23, Switzerland}
\affiliation{Department of Physics and Astronomy,
Michigan State University, East Lansing, MI 48824, USA}
\date{\today}
\begin{abstract}
Development of neutron-rich radioactive beams at the
HRIBF facility has stimulated experimental and theoretical activity in heavy Sn and Te isotopes.
Recently, the g-factor of the first 2$^+$ state in $^{132}$Te has been measured.
We report here new shell-model calculation of magnetic moments for selected Sn and Te isotopes.
The residual interaction is based on the
CD-Bonn renormalized $ G $-matrix. Single-particle spin and orbital effective g-factors
are evaluated microscopically including core polarization and meson exchange currents effects.
\end{abstract}

\maketitle

\section{Introduction}
 The $g$-factors of nuclear excited states yield valuable information on the
 make-up of their wavefunctions. The advent of radioactive beams (RIBs) constitutes a major new initiative in
 nuclear structure investigations, opening up many new
experimental opportunities. Systematics of results, obtained
with stable beams, can be extended to new areas of a special theoretical interest.
The region of neutron-rich Sn and Te nuclei with the number of protons at or just above the Z=50
closed shell is now open to direct study using Coulomb excitation in inverse kinematics.
The combination of a relative weak beam, low Coulomb excitation cross-section
and rather high energy of the low-lying excited states, implies that only
the first 2$^+$ state is currently accessible to experiment.
The B(E2; 0$^+\rightarrow$2$^+$) values for $^{132,134,136}$Te 
were measured \cite{rad02} using this technique. The results are in agreement
with systematics and model calculations for $^{132,134}$Te (N=80,82). A 
rather unexpectedly low B(E2) value was obtained for $^{136}$Te.
This anomaly is accompanied with a significant drop of 2$^+$ state energies in
the $N=84$ nuclei $^{134}$Sn and $^{136}$Te compared to the 
$N=80$ nuclei  $^{130}$Sn and $^{132}$Te.  
In this way, both the first 2$^+$ states and the $B(E2)$ values in N=80 and N=84 Sn and Te  
isotones are asymmetric with respect to $N=82$. 
The $g$-factor of the first 2$^+$ state in $^{132}$ Te has been reported recently \cite{ENAM}.

Previous shell-model calculations (see Ref.~\cite{rad02} for details) provided reasonable agreement
with energy spectra and B(E2) in N=80 and N=82 Sn and Te isotones but failed
to explain the B(E2) value in $^{136}$Te. Magnetic moments were calculated
for $^{134}$Te, $^{136,137}$Xe and $^{137}$Cs by  Sakar and Sakar \cite{sar01} with the
KH5082 and CW5082 interactions fitted in the $^{208}$Pb and scaled to the $^{132}$Sn region,
and with empirical effective single-particle $g$-factors.
Shell-model calculations for the 2$^+$, 4$^+$ and 6$^+$
states in $^{130-134}$Te and $^{132-136}$Xe were reported in Ref.~\cite{jak02},
where
the surface delta interaction (SDI) was used with two different sets of parameters.
The single particle states were chosen to reproduce single proton states in ${^{133}_{51}}$Sb  
and single neutron states in ${^{131}_{50}}$Sn. The single-particle spin and
orbital effective $g$-factors were based on the experimental $g$-factors
of the low-lying (7/2)$^+$ and (5/2)$^+$ states in the odd-$Z$, $N=82$ isotones.  

Properties of 2$^+$ states around $^{132}$Sn have been also studied by Terasaki {\em et al.} \cite{ter02}  
in a separable quadrupole-plus-pairing model. They investigated the nature and
single-particle structure of the 2$^+$ states and calculated B(E2, $0^+ \rightarrow 2^+$) values.
The $g$-factors were presented for $^{134,136,138}$Xe and $^{132,134,136}$Te.    
Single-particle bare orbital $g$-factors and bare spin $g$-factors,
multiplied by 0.7, were used in the calculation.

The present shell-model calculations  were carried out in the proton-neutron   
formalism starting with $^{132}$Sn as a closed core.
A realistic two-body residual interaction based on the CD-Bonn interaction model  \cite{cdbonn} 
was used. The magnetic moments take into account microscopic calculations
of core-polarization and meson-exchange current effects.

\section{Shell-model Hamiltonian}

The wave functions for $N \leq 82$ were
obtained in the model space of
($  0g_{7/2},1d_{5/2},1d_{3/2},2s_{1/2},0h_{11/2})^{Z-50}  $ for
proton particles and
($  0g_{7/2},1d_{5/2},1d_{3/2},2s_{1/2},0h_{11/2})^{N-82}  $ for
neutron holes.
Based upon
the energy levels observed in $^{133}$Sb \cite{San98}
the proton single-particle energies are $-$9.68, $-$8.72, $-$7.24,
 and $-$6.88 MeV for the
$  0g_{7/2}  $, $ 1d_{5/2} $ ,  $ 1d_{3/2} $    and
$  0h_{11/2}  $
orbitals, respectively. The proton $  2s_{1/2}  $ level
is not yet observed and we use an estimated energy of
$-$7.34 MeV.
Based upon
the energy levels observed in $^{131}$Sn \cite{Fire,fogel}
the neutron single-particle energies are $-$9.74, $-$8.97, $-$7.31,
$-$7.62  and $-$7.38 MeV for the
$  0g_{7/2}  $, $ 1d_{5/2} $  ,   $ 1d_{3/2} $ ,   $ 2s_{1/2} $   and
 $  0h_{11/2}  $
orbitals, respectively. Note the energy of the 11/2$^-$ state
given at 242 keV in \cite{Fire} has now been corrected to 65 keV
in \cite{fogel}.
The wave functions for $N \geq 82$ were
obtained with the same model space for protons as above and with a model
space for neutrons of
($  0h_{9/2},1f_{7/2},1f_{5/2},2p_{3/2},2p_{1/2},0i_{13/2})^{N-82}  $
with respective single-particle energies of -0.894, -2.455, -0.450,
-1.601, -0.799 and 0.25 MeV.

We used the shell-model code OXBASH \cite{OXBASH}.
The residual two-body interaction is obtained starting with a
$  G  $-matrix derived from the CD-Bonn \cite{cdbonn} nucleon-nucleon
interaction. The harmonic oscillator
basis was employed for the single-particle radial wave functions with an
oscillator energy $\hbar\omega$=7.87 MeV. The effective interaction
for the above shell-model space is obtained from
the $  \hat{Q}  $-box method and
includes all non-folded diagrams through third-order in the
interaction $  G  $ and sums up the folded diagrams to infinite
order \cite{hko95}. The Coulomb interaction was added to
the interaction between protons.
There are three parts to the Hamiltonian which will be considered
in turn, the proton-proton ($  pp  $),
neutron-neutron ($  nn  $) and proton-neutron ($  pn  $)
interactions.

In \cite{Ho97} and \cite{Su98} the $  pp  $ Hamiltonian
based upon the Bonn-A $  G  $-matrix was used to study the
N=82 isotones. The energy levels obtained with this
Hamiltonian agreed with experiment to about a hundred keV.
The present results for $^{134}$Te with CD-Bonn are shown in Table~\ref{tab1};
agreement with experiment is comparable to the previous results.
\begingroup
\begin{table}
 \caption{\label{tab1} Experimental energy levels up to 3 MeV for
$^{134}$Te compared to
the calculation with CD-Bonn.}
\begin{ruledtabular}
\begin{tabular}{rrr}
J$^{ \pi }$ & Experiment & CD-Bonn  \\
\hline \\
 0$^{ + }_{1}$ & 0.0  & 0.0    \\
 2$^{ + }_{1}$ & 1.28 & 1.21   \\
 4$^{ + }_{1}$ & 1.57 & 1.48 \\
 6$^{ + }_{1}$ & 1.69 & 1.61 \\
 6$^{ + }_{2}$ & 2.40 & 2.17  \\
 2$^{ + }_{2}$ & 2.46 & 2.45  \\
 4$^{ + }_{2}$ & 2.55 & 2.45  \\
 1$^{ + }_{1}$ & 2.63 & 2.41   \\
 3$^{ + }_{1}$ & 2.68 & 2.54  \\
 5$^{ + }_{1}$ & 2.73 & 2.54  \\
 0$^{ + }_{2}$ &      & 2.66  \\
 3$^{ + }_{3}$ & 2.93 & 3.06  \\
\end{tabular}
\end{ruledtabular}
\end{table}
\endgroup

In \cite{Ho98} the $  nn  $ Hamiltonian
based upon the CD-Bonn $  G  $-matrix was used to study the
the Sn isotones. Our results for $^{130}$Sn were similar
to these previous calculations, but differ due to the
new energy for the $h_{11/2}$ level in $^{131}$Sn \cite{fogel}.
The results for $^{130}$Sn can be improved a little by
multiplying the $nn$ renormalized  $G$-matrix by a factor of
0.90 - the spectrum for this adjusted interaction is shown
in Table~\ref{tab2}. An adjustment of this magnitude is
not unreasonable given that oscillator radial wavefunctions
were used in the calculation of the $G$-matrix elements.
\begingroup
\begin{table}
 \caption{\label{tab2}
Experimental energy levels up to 3 MeV for $^{130}$Sn compared to
the calculation with CD-Bonn.}
\begin{ruledtabular}
\begin{tabular}{rrr}
J$^{ \pi }$ & Experiment & CD-Bonn  \\
\hline \\
  0$^{ + }_{1}$ & 0.0  & 0.0    \\
  2$^{ + }_{1}$ & 1.22 & 1.38   \\
  7$^{-}_{1}$ & 1.95   & 1.87   \\
  0$^{ + }_{2}$ &      & 1.92    \\
 4$^{ + }_{1}$ & 2.00  & 2.08   \\
  2$^{ + }_{2}$ & 2.03 & 2.00   \\
 5$^{-}_{1}$ & 2.08    & 2.02   \\
  4$^{ + }_{2}$ &      & 2.08    \\
 4$^{-}_{1}$ & 2.21    & 2.12   \\
  6$^{ + }_{1}$ & 2.26 & 2.28   \\
  0$^{ + }_{3}$ &      & 2.34    \\
  8$^{ + }_{1}$ & 2.34 & 2.36  \\
  6$^{ - }_{1}$ &      & 2.36   \\
 10$^{ + }_{1}$ & 2.43 & 2.42   \\
  2$^{ + }_{3}$ &      & 2.41   \\
  5$^{ - }_{1}$ &      & 2.57   \\
  1$^{ + }_{1}$ &      & 2.58   \\
  6$^{ - }_{2}$ &      & 2.62   \\
  3$^-$,4$^+$             & 2.49 & \\
    4,5        & 2.49 & \\
               & 2.60 & \\
\end{tabular}
\end{ruledtabular}
\end{table}
\endgroup

Finally we consider the $  pn  $ Hamiltonian.
The key nucleus
in this regard is $^{132}$Sb whose spectrum is determined entirely
by the $  pn  $ interaction (together with the proton and neutron
single-particle energies).
Compared to the
$  pp  $ and $  nn  $ Hamiltonians there has been
little previous study of this interaction.
In \cite{Mach95} Mach {\em et al.} report on results using a finite 
range effective interaction based upon \cite{Ero94}.
The results with the CD-Bonn interaction are compared with
experiment in Table~\ref{tab3}. The levels given in this table
are a selected set of those observed in $\beta$-decay \cite{Stone89}
and high-spin $\gamma$-decay \cite{Daley}. The levels are    
labeled by their dominant theoretical component, although
typically they are only about $80\%$ pure.   
\begingroup
\begin{table}
 \caption{\label{tab3}
Experimental energy levels (in MeV) for $^{132}$Sb compared to
the calculation with CD-Bonn.
Energies in square brackets are relative to that for the 8$^{-}_{1}$
levels that is estimated to be near 0.20 MeV.
}
\begin{ruledtabular}
\begin{tabular}{rrrr}
J$^{ \pi }$ & Experiment & CD-Bonn & Main Conf\\
\hline \\
 4$^{ + }_{1}$ &     0 &     0 & $\pi$g$_{7/2}$-$\nu$d$_{3/2}$
\\
 3$^{ + }_{1}$ &   0.08 &   0.12 & \\
 5$^{ + }_{1}$ &   0.16 &   0.26 & \\
 2$^{ + }_{1}$ &   0.43 &   0.52 & \\
\hline \\
 3$^{ + }_{2}$ &   0.53 &   0.56 & $\pi$g$_{7/2}$-$\nu$s$_{1/2}$
\\
\hline \\
 1$^{ + }_{1}$ &  1.32 &  1.56 & $\pi$d$_{5/2}$-$\nu$d$_{3/2}$
\\
 1$^{ + }_{2}$ &  2.27 &  2.27 & $\pi$d$_{7/2}$-$\nu$d$_{5/2}$
\\
\hline \\
 (8$^{-}_{1}$) & [0.20] &   0.24 &
$\pi$g$_{7/2}$-$\nu$h$_{11/2}$ \\
 (6$^{-}_{1}$) & 0.25 &   0.32 &   \\
 (5$^{-}_{1}$) & 0.39 &   0.35 &   \\
 (4$^{-}_{1}$) & 0.48 &   0.47 &  \\
 (9$^{-}_{1}$) & [1.22] &  0.84 &  \\
\hline \\
 (10$^{ + }_{1}$) & [3.00] &  2.93 &
$\pi$h$_{11/2}$-$\nu$h$_{11/2}$ \\
 (11$^{ + }_{1}$) & [3.40] &  3.28 &
$\pi$h$_{11/2}$-$\nu$h$_{11/2}$ \\
\end{tabular}
\end{ruledtabular}
\end{table}
\endgroup

Andreozzi {\em et al.} \cite{And98} have reported calculations     
using the Bonn-A $  G  $-matrix. Their method, however, differs
in several respects from the one we use. They use isospin
formalism with a $^{100}$Sn closed shell in contrast to
our proton-neutron formalism starting with $^{132}$Sn
as a closed shell core.
The main drawback of the isospin formalism is that
proton and neutron single-particle energies are not
independent and do not reproduce the experimental
levels of $^{133}$Sb and $^{131}$Sn exactly. Another
aspect is that the renormalization of the $  G  $-matrix is
carried out only to second order. The result of
including the 3rd order terms as we do is to make the
$  pn  $ matrix elements somewhat larger.
As also demonstrated in \cite{hko95},
the total interaction
in the $  T=0  $ channel to third order is of the
order of $  20\%  $ compared  with the interaction
up to second order, whereas in the $  T=1  $ channel the difference
is less than $  10\%  $. This is mainly due to the
influence of the strong nuclear tensor force
component stemming from the $  ^{3}S_{1}-^{3}D_{1}  $ partial wave
in the nucleon-nucleon interaction.

Finally as a test of the configuration mixing involving all components
of the hamiltonian, we show the calculated levels for $^{132}$Te
compared to experiment in Table~\ref{tabte134}. Experiment is taken from
the new results of Hughes {\em et al.} \cite{hughes}. The agreement between 
experiment and calculation is excellent. The 2$^+$ states are in significantly better
agreement with experiment than those given in the
separable quadrupole-plus-pairing model of Terasaki {\em et al.}                  
\cite{ter02,hughes}. In \cite{hughes} 2$^+$ assignments for the 2.25 and 2.36 MeV
states were made on the assumption that there should not be any 1$^+$
states at this low excitation energy. However, our calculation predicts
two 1$^+$ states near 2.4 MeV, and on this basis the states
observed in experiment should be labeled (1,2)$^+$. The dominant
component of these 1$^+$ states are related to the
low-lying 1$^+$ states in $^{134}$Te and $^{130}$Sn.
With regard to the
magnetic moment of the 2$^+$ state discussed here we give the components
of this wavefunction (those with probabilities of greater than one percent)
in Table~\ref{tabte134w}. We can also decompose the 2$^+$ wavefunction
in terms of the coupling between the $^{134}$Te two-proton
configuration $pp$ and the $^{130}$Sn two-neutron hole configuration $nn$.
This coupling is dominated by two components, 48.9\% for $pp(2^+)nn(0^+)$
and 32.1\% for $pp(0^+)nn(2^+)$.
\begingroup
\begin{table}
 \caption{\label{tabte134}
Experimental energy levels up to 2.5 MeV for $^{132}$Te compared to
the calculation with CD-Bonn.}
\begin{ruledtabular}
\begin{tabular}{rrr}
J$^{ \pi }$ & Experiment & CD-Bonn  \\
\hline \\
  0$^{ + }_{1}$ & 0.0      & 0.0    \\
  2$^{ + }_{1}$ & 0.97     & 0.95    \\
  (2)$^{ + }_{2}$ & 1.66   & 1.64   \\
  4$^{ + }_{1}$ & 1.67     & 1.54   \\
  6$^{ + }_{1}$ & 1.77     & 1.68   \\
  0$^{ + }_{2}$ &          & 1.70   \\
  (2)$^{ + }_{3}$ & 1.79   & 1.93   \\
  (7)$^{ - }_{1}$ & 1.92   & 1.88   \\
  (5)$^{ - }_{1}$ & 2.05   & 2.01   \\
  4$^{ - }_{1}$ &          & 2.12   \\
  0$^{ + }_{3}$ &          & 2.17   \\
  4$^{ + }_{2}$ &          & 2.20   \\
  6$^{ + }_{2}$ &          & 2.21   \\
  (2)$^{ + }_{4}$ & 2.25   & 2.25  \\
  1$^{ + }_{1}$ &          & 2.36   \\
  1$^{ + }_{2}$ &          & 2.41   \\
  6$^{ + }_{1}$ &          & 2.43   \\
  (2)$^{ + }_{5}$ & 2.36   & 2.46   \\
  4$^{ + }_{3}$ &          & 2.48   \\
\end{tabular}
\end{ruledtabular}
\end{table}
\endgroup

\begingroup
\begin{table*}
 \caption{\label{tabte134w}
Wavefunction components for the first 2$^+$ state in
$^{132}$Te (those greater than one percent).}
\begin{ruledtabular}
\begin{tabular}{ccc}
proton wavefunction & neutron wavefunction & probability  \\
\hline \\
$  (0g_{7/2})^2,(1d_{5/2})^0  $ &
$  (0g_{7/2})^8,(1d_{5/2})^6,(1d_{3/2})^2,(2s_{1/2})^2,(0h_{11/2})^{12} $ &
28.4 \\
$  (0g_{7/2})^2,(1d_{5/2})^0  $ &
$  (0g_{7/2})^8,(1d_{5/2})^6,(1d_{3/2})^4,(2s_{1/2})^2,(0h_{11/2})^{10} $ &
21.0 \\
$  (0g_{7/2})^2,(1d_{5/2})^0  $ &
$  (0g_{7/2})^8,(1d_{5/2})^6,(1d_{3/2})^3,(2s_{1/2})^1,(0h_{11/2})^{12} $ &
15.3 \\
$  (0g_{7/2})^2,(1d_{5/2})^0  $ &
$  (0g_{7/2})^8,(1d_{5/2})^6,(1d_{3/2})^4,(2s_{1/2})^0,(0h_{11/2})^{12} $ &
8.2 \\
$  (0g_{7/2})^2,(1d_{5/2})^0  $ &
$  (0g_{7/2})^8,(1d_{5/2})^4,(1d_{3/2})^4,(2s_{1/2})^2,(0h_{11/2})^{12} $ &
5.5 \\
$  (0g_{7/2})^2,(1d_{5/2})^0  $ &
$  (0g_{7/2})^8,(1d_{5/2})^5,(1d_{3/2})^4,(2s_{1/2})^1,(0h_{11/2})^{12} $ &
4.3 \\
$  (0g_{7/2})^2,(1d_{5/2})^0  $ &
$  (0g_{7/2})^6,(1d_{5/2})^6,(1d_{3/2})^4,(2s_{1/2})^2,(0h_{11/2})^{12} $ &
3.1 \\
$  (0g_{7/2})^2,(1d_{5/2})^0  $ &
$  (0g_{7/2})^7,(1d_{5/2})^6,(1d_{3/2})^3,(2s_{1/2})^2,(0h_{11/2})^{12} $ &
2.8 \\
$  (0g_{7/2})^2,(1d_{5/2})^0  $ &
$  (0g_{7/2})^8,(1d_{5/2})^5,(1d_{3/2})^3,(2s_{1/2})^2,(0h_{11/2})^{12} $ &
1.8 \\
$  (0g_{7/2})^0,(1d_{5/2})^2  $ &
$  (0g_{7/2})^8,(1d_{5/2})^6,(1d_{3/2})^4,(2s_{1/2})^2,(0h_{11/2})^{10} $ &
1.1 \\
\end{tabular}
\end{ruledtabular}
\end{table*}
\endgroup

\section{Magnetic moments}

The magnetic moment matrix element, expressed in terms of
a reduced matrix element using the Wigner-Eckart theorem       
for an operator of rank $\lambda=1$, is
\begin{widetext}
\begin{equation}
<\omega J M=J\mid \hat{\mu} \mid \omega J M=J>\, =
 \tj  {J}  {\lambda}   {J'}  {-J}  {0}   {J}
 <\omega J||\hat{\mu}||\omega J>
\end{equation}
\end{widetext}
The many-body reduced matrix element
can be expressed as a sum of products over one-body transition densities
(OBTD) times reduced single-particle matrix elements
\begin{equation}
<\omega  J||\hat{\mu}||\omega J>
= \displaystyle\sum _{k_{\alpha } k_{\beta }} {\rm OBTD}(\omega J
k_{\alpha} k_{\beta} \lambda )
 <k_{\alpha }||\mu||k_{\beta }>,
\end{equation}
where the OBTD is given by
\begin{equation}
{\rm OBTD}(\omega J k_{\alpha } k_{\beta } \lambda )
= { <\omega  J||[a^{+}_{k_{\alpha }}\otimes
\tilde{a}_{k_{\beta}}]^{\lambda }|| \omega J>
\over  \sqrt{(2\lambda+1)}}.
\end{equation}
The sum is over all pairs of orbits for protons and neutrons
that can couple up to a tensor of rank $\lambda=1$.

The free-nucleon operator is defined as
\begin{equation}
\label{eq1f}
{\bf \mu}_{{\rm free}} = g_{l} {\bf l} +
g_{s} {\bf s},
\end{equation}
with $  g_{l}({\rm proton}) = 1.0  $,
$  g_{l}({\rm neutron}) = 0.0  $,
$  g_{s}({\rm proton}) = 5.587  $,
$  g_{s}({\rm neutron}) = -3.826  $.

The magnetic moment operator in finite nuclei is modified from the
free-nucleon operator
due to core-polarization and meson-exchange
current (MEC) corrections \cite{To87,CT90}.
The effective operator is defined as
\begin{equation}
\label{eq1}
{\bf \mu}_{{\rm eff}} = g_{l,{\rm eff}} {\bf l} +
g_{s,{\rm eff}} {\bf s} +
g_{p,{\rm eff}} [Y_{2} ,{\bf s}] ,
\end{equation}
\noindent where $  g_{x,{\rm eff}} = g_{x} + \delta g_{x}  $, $  x = l
$,
$  s  $ or $  p  $,
with $  g_{x}  $ the free-nucleon, single-particle
$  g  $-factors ($g_p=0$) and $  \delta g_{x}  $ the
calculated correction to it.  Note the presence of a new term
$  [Y_{2} , {\bf s}]  $, absent from the free-nucleon operator, which is
a spherical
harmonic of rank $\lambda'=2$ coupled to a spin operator to form a spherical tensor
of multipolarity $\lambda=1$.

The corrections, $  \delta g_{x}  $, are computed in perturbation
theory for
the closed-shell-plus-or-minus-one configuration with the closed shell
being $^{132}$Sn.  The first-order core-polarization correction
involves coupling the valence nucleon to the 1$^{ + }$ particle-hole
states: proton $  (0g_{9/2}^{-1},0g_{7/2})  $ and neutron
$  (0h_{11/2}^{-1},0h_{9/2})  $.  This term leads to a large quenching
in the
$  g_{s,{\rm eff}}  $ value but only a small change in
$  g_{l,{\rm eff}}  $.  The calculation is easily extended to all orders
in the RPA series, \cite{To87}.  The residual interaction in these
calculations is taken as a one-boson-exchange potential multiplied by
a short-range correlation function.  This modification is an
approximate, but easy, way to obtain a $  G  $-matrix.

Meson-exchange current corrections arise because nucleons in nuclei are
interacting through the exchange of mesons, which can be disturbed
by the electromagnetic field.  Since meson exchange involves two
nucleons, the correction leads to two-body magnetic moment operators.
In a closed-shell-plus-or-minus-one configuration, computation of this
correction requires evaluation of the two-body matrix elements between
the valence nucleon and one of the core nucleons, summed over all
nucleons in the core.  The results can be expressed in terms of an
equivalent effective one-body operator, Eq.~(\ref{eq1}), acting on   
the valence nucleon alone.  The details of the two-body MEC operators
are described in \cite{To87} and updated in \cite{To96}.  For
consistency, the same mesons, coupling constants, masses and
short-range correlations are used in the construction of the MEC
operators
as are used in the one-boson-exchange potential.

There are two further terms to consider.  First is a mesonic correction
in which the meson prompts the nucleon to be raised to an excited
state, the $  \Delta  $-isobar resonance, which is then de-excited by
the
electromagnetic field.  This correction leads to a two-body operator
that is handled like the MEC correction.  Second is a relativistic
correction to the one-body operator, \cite{To87}.  Both these
corrections
amount to only a few percent change to the magnetic moment, but are
retained for completeness.

Finally there are other second-order core-polarization corrections not
contained in the RPA series that are difficult to compute because there
are
no selection rules to limit the number of intermediate states to be
summed.  A further correction of the same order in meson-nucleon
couplings
is a core-polarization correction to the two-body MEC operator.
Fortunately, as Arima {\it et al.} \cite{AH79,HAS80} have
pointed out,
the latter terms largely cancel the former.
In our earlier work \cite{St97}
this correction was not explicitly calculated, but effective $  g
$-factors
from a comparable calculation in Pb were used.  Here we have computed
these terms, so our result differs a little from \cite{St97} but not
significantly.  The computation, however, was performed approximately.
The closed shell was taken to be an $LS$-closed shell, with $A=140$,   
and the computation performed in $LS$-coupling.  This leads to a great   
saving in computation time and makes the calculation tractable
However, the neutron excess orbitals are not now treated correctly.
The intermediate-state summation is explicitly computed up
$  12 \hbar \omega  $ and geometrically extrapolated beyond that.

The resulting corrections to the $  g  $-factors from the sum of all
these effects are listed in Table~\ref{tab6}. 
It is evident that there is not a great 
deal of state dependence in the effective operator.  Thus for orbitals not 
explicitly listed in Table VI, we have used average values for their 
effective $g$-factors:  protons: $\delta g_l=0.094$, 
$\delta g_s=-2.14$, $\delta g_p=2.03$, while for neutrons:
$\delta g_l=-0.039$, $\delta g_s=1.92$, and $\delta g_p =-0.93$.
\begingroup
\begin{table*}
 \caption{\label{tab6}Effective $  g  $-factors from core-polarization and MEC calculations.}
\begin{ruledtabular}
\begin{tabular}{crrrcrrr}
 & \multicolumn{3}{c}{Proton} & & \multicolumn{3}{c}{Neutron} \\
\cline{2-4}
\cline{6-8}
 & & & & & & & \\ [-3mm]
Orbital & $  \delta g_{l}   $ & $  \delta g_{s}   $ & $  \delta g_{p}
$ & &
$  \delta g_{l}   $ & $  \delta g_{s}   $  & $  \delta g_{p}   $ \\
\hline \\ [-3mm]
   $  0h  $  &    0.087 & $-$1.988 &    1.549 & &
            $-$0.033  &  1.847 & $-$0.769 \\
  $  0g  $  &   0.131 & $-$2.284  &  1.705 & &
            $-$0.067 &   2.033 & $-$0.591  \\
  $  1d  $  &   0.063 & $-$2.167 &   1.681  & &
            $-$0.018  &  1.976 &  $-$1.039  \\
  $  2s  $  &  &         $-$2.102  & & &
                &  1.804  & \\
 $  0g-1d  $  & & &                   3.329 & &
          & &            $-$1.162 \\
 $  1d-2s  $  & & &                   1.902 & &
          & &            $-$1.093 \\
\end{tabular}
\end{ruledtabular}
\end{table*}
\endgroup
All matrix elements have
been evaluated with harmonic oscillator radial functions of
characteristic frequency $  \hbar \omega = 7.87  $ MeV.  Note that with
a term $  [Y_{2}, {\bf s}]  $, in the effective magnetic moment
operator,
Eq.~\ref{eq1}, there are non-zero off-diagonal matrix elements
between $  0g_{7/2}-1d_{5/2}  $
and $  1d_{3/2}-2s_{1/2}  $ orbitals.  These $  l  $-forbidden matrix
elements
are zero with the free-nucleon
operator but non-zero here.  However, their impact in the present
calculation is very small.
The results in Table~\ref{tab6} strictly only apply to
closed-shell-plus-or-minus-one configurations at a $^{132}$Sn closed
shell. An occupation number-dependent effective operator was
introduced in \cite{white} to account for the effects of blocking in the
core-polarization.  However, for the nuclei considered here
the occupation number dependence for the magnetic moments
moments is not large (on the order of 0.02), and we do not
include this effect.

The magnetic moment for the 7/2$^+$ ground state of the
$N=82$ nucleus $^{133}$Sb is 3.00(1) \cite{St97} compared to the 
free-nucleon value of 1.717 and effective operator
value of 2.925. Thus, as discussed in \cite{St97}, the
core-polarization and mesonic exchange corrections 
are essential for understanding the enhancement of this
magnetic moment relative to its free-nucleon (Schmidt)
value. Results discussed in \cite{white} for other
$N=82$ nuclei $^{135}$I, $^{137}$Cs and $^{139}$La
show the general importance of the effective operator
for these more complicated configurations. 

Experimental magnetic moments for low-lying excitations
in a range of even-even isotopes close to $^{132}$Sn are
compared in Table~\ref{tab7} to those
obtained with the free-nucleon $  g  $-factors.
\begingroup
\begin{table*}
 \caption{\label{tab7}Experimental and calculated magnetic moments.
The calculations use the free-nucleon $  g  $-factors and
 effective $  g  $-factors. The last two columns give the proton
and neutron contributions 
to the effective operator moments. Results for 4$^{ + }$ and 6$^{ + }$ states
are added
for comparison to related experiment data.}
\begin{ruledtabular}
\begin{tabular}{rrrrrrrr}
 Nuclide & $n$ & J$^{ \pi }$ & Experiment  &
Effective & Free & proton & neutron \\
\hline
 $^{124}$Sn & 74 & 2$^{ + }$ & $-$0.3(2)\protect\cite{njs}  & $-$0.270 & $-$0.364 & 0 & $-$0.270 \\
 $^{126}$Sn & 76 & 2$^{ + }$ &          & $-$0.262  & $-$0.355 & 0 & $-$0.262  \\
 $^{128}$Sn & 78 & 2$^{ + }$ &          & $-$0.253  & $-$0.343 & 0 & $-$0.253   \\
 $^{130}$Sn & 80 & 2$^{ + }$ &          & $-$0.275  & $-$0.385 & 0 & $-$0.275   \\
 $^{134}$Sn & 84 & 2$^{ + }$ &          & $-$0.469  & $-$0.745 & 0 & $-$0.469\\
\hline
 $^{130}$Te & 78 & 2$^{ + }$ & 0.59(7)\protect\cite{jak02}  & 0.693 & 0.360  & 0.806 & $-$0.113  \\
 $^{132}$Te & 80 & 2$^{ + }$ & 0.70(10)\protect\cite{ENAM} & 0.975 & 0.575  & 1.027 & $-$0.052\\
 $^{132}$Te & 80 & 4$^{ + }$ &          & 3.18   & 1.90  & 3.20 & $-$0.02  \\
 $^{132}$Te & 80 & 6$^{ + }$ & 5.08(15)\protect\cite{jak02} & 5.14   & 3.20  & 5.15 & $-$0.01  \\
 $^{134}$Te & 82 & 2$^{ + }$ &          & 1.724 & 1.035  & 1.724 & 0 \\
 $^{134}$Te & 82 & 4$^{ + }$ &          & 3.44   & 2.04  & 3.44 & 0 \\
 $^{134}$Te & 82 & 6$^{ + }$ & 4.7(6)\protect\cite{jak02}   & 5.20   & 3.15  & 5.20 & 0 \\
 $^{136}$Te & 84 & 2$^{ + }$ &          & 0.695 & 0.544  & 0.846 & $-$0.151\\
\hline
 $^{134}$Xe & 80 & 2$^{ + }$ & 0.708(14)\protect\cite{jak02}  & 0.825  & 0.541  & 0.886 & $-$0.061 \\
 $^{136}$Xe & 82 & 4$^{ + }$ & 3.2(6)\protect\cite{njs}   & 3.55   & 2.17  & 3.55 & 0 \\
 $^{136}$Xe & 82 & 2$^{ + }$ & 1.53(9)\protect\cite{jak02}  & 1.823  & 1.165  & 1.823 & 0\\
 $^{138}$Xe & 84 & 2$^{ + }$ &          & 0.775  & 0.623 & 0.912 & $-$0.137 \\
 $^{138}$Ba & 82 & 2$^{ + }$ & 1.44(22)\protect\cite{njs} & 2.00   & 1.52  & 2.00 & 0 \\
\end{tabular}
\end{ruledtabular}
\end{table*}
\endgroup

\section{Discussion and conclusions}

Examination of Table~\ref{tab7} illustrates the quality of the present shell-model calculation.
We stress that all the magnetic dipole moments are calculated in full
model space and the effective $g$-factors are obtained from microscopic calculation.
In comparison with the QRPA model of Terasaki {\em et al.} \cite{ter02} we do not predict a
dramatic decrease and change in sign of the magnetic moment of the 2$^+_1$ state in $^{136}$Te
which might have risen in that model as a consequence of overestimation of the contribution
of neutron excitations to the total wavefunction. The experimental determination
of this $g$-factor is clearly of great interest \cite{isoldete136}. 
More generally, we find the
contribution of the neutron components to magnetic dipole moment of low-lying
states in Te, Xe and Ba isotopes much smaller than in Sn nuclei.

We see from Table \ref{tab7} that the calculated magnetic moments for the 2$^+$ states
have a maximum at $N=82$ for the pure proton configuration.
The main deviation between experiment and theory can be traced to the magnetic
moments of the proton $2^+$ states for $N=82$. The calculated moment with the effective operator for
the $^{136}$Xe 2$^{ + }$ is about 15 percent larger than experiment. For $N=80$ 
if we reduce the
proton contribution by 15 percent we would get 0.82 for the $^{132}$Te 2$^{ + }$ moment
and 0.69 for the $^{134}$Xe 2$^{ + }$ moment, in better
agreement with the experimental values of 0.70(10) and 0.708(14), respectively. 

In comparison with the previous shell-model calculations,
we note that in the calculations of Jakob et al. \cite{jak02} (the same model space as ours),
the schematic SDI interaction was fitted to a set of single particle energies
containing the old value of the 11/2$^-$ state in $^{131}$Sn which is $\sim$200 keV
higher than the new value \cite{fogel}. The levels schemes obtained with the SDI interaction
are not shown in \cite{jak02}. The effective magnetic moment operator in \cite{jak02}
was obtained from a fit of $g$-factors in odd-even nuclei.
The values obtained are close to our microscopic results, but the fitted operator does
not include the tensor-type correction obtained microscopically. The calculations
of Sarkar and Sarkar \cite{sar01} are 
based on a fitted interaction for N=82 plus a
renormalized $ G $- matrix extrapolated from the region of $^{208}$Pb for N$>$82. Their wavefunctions
for N=82 
should be comparable in accuracy to the present model. Again however, 
an effective magnetic moment
operator used in the 
evaluation of magnetic moments is fitted to
experimental data and does not include the tensor-type term.

In summary, our microscopic interaction is based on a realistic nucleon-nucleon
interaction and yields an excellent agreement with experiment for the energy levels
of nuclei near $^{132}$Sn. This is the first fully microscopic calculation,
using effective $g$-factors obtained from calculated core-polarization and mesonic exchange corrections.
The good agreement with experiment confirms not only the validity of the shell
model hamiltonian but also of the microscopic effective $g$-factors.
Significant predictions are made to stimulate 
further experimental $g$-factor
measurements in this region.

\section{Acknowledgements}
This research has been supported by the US NSF grant PHY-0244453 (BAB),
US DOE grants no. DE-FG02-96ER40983(NJS), DE-FG02-94ER40834 (JRS) and the Research Council of Norway.

\end{document}